\documentclass[conference]{IEEEtran}
\usepackage[nolist]{acronym} 
\usepackage{tikz}
\usepackage{pgfplots}
\usepgfplotslibrary{groupplots}
\usepackage{graphicx}
\usepackage{subfloat}
\usepackage{subfig}
\usepackage{caption}
\usepackage{cite}
\usetikzlibrary{spy}
\usepackage{float}
\pgfplotsset{compat=newest}
\usepackage{hyperref}
\usepackage{units}
\usepackage{amsmath, amsbsy, amssymb}
\usepackage{tikzscale}
\usepackage{color}
\usepackage{epigraph}
\usepackage{circuitikz}
\usepackage{multirow}
\usetikzlibrary{shapes,arrows}
\usetikzlibrary{external}
\usepackage[ruled,boxed,lined]{algorithm2e}

\usepackage[group-separator={,}]{siunitx}
\definecolor{darkgreen}{rgb}{0.125,0.5,0.169}
\definecolor{mittelblau}{RGB}{0, 126, 198}
\definecolor{violettblau}{cmyk}{0.9, 0.6, 0, 0}
\definecolor{rot}{RGB}{238, 28 35}
\definecolor{apfelgruen}{RGB}{140, 198, 62}
\definecolor{gelb}{RGB}{1, 221, 0}
\definecolor{orange}{RGB}{244, 111, 33}
\definecolor{pink}{RGB}{237, 0, 140}
\definecolor{lila}{RGB}{128, 10, 145}
\definecolor{hellgrau}{RGB}{224, 224, 224}
\definecolor{grau}{RGB}{128, 128, 128}
\definecolor{dunkelgrau}{RGB}{80,80,80}
\definecolor{anthrazit}{RGB}{19, 31, 31}

\pgfplotscreateplotcyclelist{corporate colours markers}{%
rot, every mark/.append style={fill=.!80!rot},mark=*\\%
mittelblau, every mark/.append style={fill=.!80!mittelblau},mark=square*\\%
apfelgruen, every mark/.append style={fill=.!80!apfelgruen},mark=triangle*\\%
orange, mark=star\\%
pink, every mark/.append style={fill=.!80!pink},mark=diamond*\\%
violettblau, every mark/.append style={fill=.!80!violettblau},mark=otimes*\\%
lila, mark=|\\%
gelb, every mark/.append style={fill=.!80!gelb},mark=pentagon*\\%
hellgrau, mark=text,text mark=p\\%
anthrazit, mark=text,text mark=a\\%
}

\pgfplotscreateplotcyclelist{corporate colours markers double}{%
rot, every mark/.append style={fill=.!80!rot},mark=*\\%
rot, dashed, every mark/.append style={fill=.!80!rot,solid},mark=*\\%
mittelblau, every mark/.append style={fill=.!80!mittelblau},mark=square*\\%
mittelblau, dashed, every mark/.append style={fill=.!80!mittelblau,solid},mark=square*\\%
apfelgruen, every mark/.append style={fill=.!80!apfelgruen},mark=triangle*\\%
apfelgruen, dashed, every mark/.append style={fill=.!80!apfelgruen,solid},mark=triangle*\\%
orange, mark=star\\%
orange, dashed, mark=star\\%
pink, every mark/.append style={fill=.!80!pink},mark=diamond*\\%
pink, dashed, every mark/.append style={fill=.!80!pink,solid},mark=diamond*\\%
violettblau, every mark/.append style={fill=.!80!violettblau},mark=otimes*\\%
violettblau, dashed, every mark/.append style={fill=.!80!violettblau,solid},mark=otimes*\\%
lila, mark=|\\%
lila, dashed, mark=|\\%
gelb, every mark/.append style={fill=.!80!gelb},mark=pentagon*\\%
gelb, dashed, every mark/.append style={fill=.!80!gelb,solid},mark=pentagon*\\%
}
\usetikzlibrary{shapes,arrows}

\begin{document}

\title{Towards Practical Indoor Positioning Based on Massive MIMO Systems}
\author{\IEEEauthorblockN{Mark Widmaier, Maximilian Arnold, Sebastian D\"orner, Sebastian Cammerer, and Stephan ten Brink}
\IEEEauthorblockA{Institute of Telecommunications, University of  Stuttgart, 70659 Stuttgart, Germany 
    \\\{arnold,doerner,cammerer,tenbrink\}@inue.uni-stuttgart.de
}
}
\maketitle
% Vectors
\renewcommand{\vec}[1]{\mathbf{#1}}
\newcommand{\vecs}[1]{\boldsymbol{#1}}

\newcommand{\av}{\vec{a}}
\newcommand{\bv}{\vec{b}}
\newcommand{\cv}{\vec{c}}
\newcommand{\dv}{\vec{d}}
\newcommand{\ev}{\vec{e}}
\newcommand{\fv}{\vec{f}}
\newcommand{\gv}{\vec{g}}
\newcommand{\hv}{\vec{h}}
\newcommand{\iv}{\vec{i}}
\newcommand{\jv}{\vec{j}}
\newcommand{\kv}{\vec{k}}
\newcommand{\lv}{\vec{l}}
\newcommand{\mv}{\vec{m}}
\newcommand{\nv}{\vec{n}}
\newcommand{\ov}{\vec{o}}
\newcommand{\pv}{\vec{p}}
\newcommand{\qv}{\vec{q}}
\newcommand{\rv}{\vec{r}}
\newcommand{\sv}{\vec{s}}
\newcommand{\tv}{\vec{t}}
\newcommand{\uv}{\vec{u}}
\newcommand{\vv}{\vec{v}}
\newcommand{\wv}{\vec{w}}
\newcommand{\xv}{\vec{x}}
\newcommand{\yv}{\vec{y}}
\newcommand{\zv}{\vec{z}}
\newcommand{\zerov}{\vec{0}}
\newcommand{\onev}{\vec{1}}
\newcommand{\alphav}{\vecs{\alpha}}
\newcommand{\betav}{\vecs{\beta}}
\newcommand{\gammav}{\vecs{\gamma}}
\newcommand{\lambdav}{\vecs{\lambda}}
\newcommand{\omegav}{\vecs{\omega}}
\newcommand{\sigmav}{\vecs{\sigma}}
\newcommand{\tauv}{\vecs{\tau}}

% Matrices
\newcommand{\Am}{\vec{A}}
\newcommand{\Bm}{\vec{B}}
\newcommand{\Cm}{\vec{C}}
\newcommand{\Dm}{\vec{D}}
\newcommand{\Em}{\vec{E}}
\newcommand{\Fm}{\vec{F}}
\newcommand{\Gm}{\vec{G}}
\newcommand{\Hm}{\vec{H}}
\newcommand{\Id}{\vec{I}}
\newcommand{\Jm}{\vec{J}}
\newcommand{\Km}{\vec{K}}
\newcommand{\Lm}{\vec{L}}
\newcommand{\Mm}{\vec{M}}
\newcommand{\Nm}{\vec{N}}
\newcommand{\Om}{\vec{O}}
\newcommand{\Pm}{\vec{P}}
\newcommand{\Qm}{\vec{Q}}
\newcommand{\Rm}{\vec{R}}
\newcommand{\Sm}{\vec{S}}
\newcommand{\Tm}{\vec{T}}
\newcommand{\Um}{\vec{U}}
\newcommand{\Vm}{\vec{V}}
\newcommand{\Wm}{\vec{W}}
\newcommand{\Xm}{\vec{X}}
\newcommand{\Ym}{\vec{Y}}
\newcommand{\Zm}{\vec{Z}}
\newcommand{\Lambdam}{\vecs{\Lambda}}
\newcommand{\Pim}{\vecs{\Pi}}

% Calligraphic
\newcommand{\Ac}{{\cal A}}
\newcommand{\Bc}{{\cal B}}
\newcommand{\Cc}{{\cal C}}
\newcommand{\Dc}{{\cal D}}
\newcommand{\Ec}{{\cal E}}
\newcommand{\Fc}{{\cal F}}
\newcommand{\Gc}{{\cal G}}
\newcommand{\Hc}{{\cal H}}
\newcommand{\Ic}{{\cal I}}
\newcommand{\Jc}{{\cal J}}
\newcommand{\Kc}{{\cal K}}
\newcommand{\Lc}{{\cal L}}
\newcommand{\Mc}{{\cal M}}
\newcommand{\Nc}{{\cal N}}
\newcommand{\Oc}{{\cal O}}
\newcommand{\Pc}{{\cal P}}
\newcommand{\Qc}{{\cal Q}}
\newcommand{\Rc}{{\cal R}}
\newcommand{\Sc}{{\cal S}}
\newcommand{\Tc}{{\cal T}}
\newcommand{\Uc}{{\cal U}}
\newcommand{\Wc}{{\cal W}}
\newcommand{\Vc}{{\cal V}}
\newcommand{\Xc}{{\cal X}}
\newcommand{\Yc}{{\cal Y}}
\newcommand{\Zc}{{\cal Z}}

\newcommand{\CN}{\Cc\Nc}

% Number sets
\newcommand{\CC}{\mathbb{C}}
\newcommand{\MM}{\mathbb{M}}
\newcommand{\NN}{\mathbb{N}}
\newcommand{\RR}{\mathbb{R}}

% Mixed symbols
\newcommand{\htp}{^{\mathsf{H}}}
\newcommand{\tp}{^{\mathsf{T}}}

% Brackets
\newcommand{\LB}{\left(}
\newcommand{\RB}{\right)}
\newcommand{\LP}{\left\{}
\newcommand{\RP}{\right\}}
\newcommand{\LSB}{\left[}
\newcommand{\RSB}{\right]}

\renewcommand{\ln}[1]{\mathop{\mathrm{ln}}\LB #1\RB}
\newcommand\norm[1]{\left\lVert#1\right\rVert}
\newcommand{\cs}[1]{\mathop{\mathrm{cs}}\LSB #1\RSB}

% Expectation, Variance, etc
\newcommand{\EE}{{\mathbb{E}}}
\newcommand{\Expect}[2]{\EE_{#1}\LSB #2\RSB}

% Theorems, Lemma, etc.
\newtheorem{definition}{Definition}[section]
\newtheorem{remark}{Remark}

\begin{acronym}
 \acro{CSI}{channel state information}
 \acro{UE}{user equipment}
 \acro{UL}{uplink}
 \acro{BS}{base station}
 \acro{TDD}{time division duplexing}
 \acro{FDD}{frequency division duplexing}
 \acro{ECC}{error-correcting code}
 \acro{MLD}{maximum likelihood decoding}
 \acro{HDD}{hard decision decoding}
 \acro{IF}{intermediate frequency}
 \acro{RF}{radio frequency}
 \acro{SDD}{soft decision decoding}
 \acro{NND}{neural network decoding}
 \acro{CNN}{convolutional neural network}
 \acro{ML}{maximum likelihood}
 \acro{GPU}{graphical processing unit}
 \acro{BP}{belief propagation}
 \acro{LTE}{Long Term Evolution}
 \acro{BER}{bit error rate}
 \acro{SNR}{signal-to-noise-ratio}
 \acro{ReLU}{rectified linear unit}
 \acro{BPSK}{binary phase shift keying}
 \acro{QPSK}{quadrature phase shift keying}
 \acro{AWGN}{additive white Gaussian noise}
 \acro{MSE}{mean squared error}
 \acro{MDE}{mean distance error}
 \acro{MDA}{mean distance accuracy}
 \acro{FT}{fine-tuning}
 \acro{IoT}{Internet of Things}
 \acro{LLR}{log-likelihood ratio}
 \acro{MAP}{maximum a posteriori}
 \acro{NVE}{normalized validation error}
 \acro{BCE}{binary cross-entropy}
 \acro{BLER}{block error rate}
 \acro{SQR}{signal-to-quantisation-noise-ratio}
 \acro{MIMO}{multiple input multiple output}
 \acro{OFDM}{orthogonal frequency division multiplex}
 \acro{RF}{radio frequency}
 \acro{LOS}{line of sight}
 \acro{NLOS}{non-line of sight}
 \acro{NMSE}{normalized mean squared error}
 \acro{CFO}{carrier frequency offset}
 \acro{GPS}{global positioning system}
 \acro{DE}{distance error}
 \acro{DS}{delay spread}
 \acro{SFO}{sampling frequency offset}
 \acro{IPS}{indoor positioning system}
 \acro{OPS}{outdoor positioning system}
 \acro{IoT}{internet of things} 
 \acro{TRIPS}{time-reversal IPS}
 \acro{RSSI}{received signal strength indicator}
 \acro{MIMO}{multiple-input multiple-output}
 \acro{ENoB}{effective number of bits}
 \acro{AGC}{automated gain control}
 \acro{ADC}{analog to digital converter}
 \acro{ADCs}{analog to digital converters}
 \acro{FB}{front bandpass}
 \acro{FPGA}{field programmable gate array}
 \acro{JSDM}{Joint Spatial Division and Multiplexing}
 \acro{NN}{Neural Network}
 \acro{IF}{intermediate frequency}
 \acro{LoS}{Line-of-Sight}
 \acro{RNN}{recurrent neural network}
 \acro{NLoS}{Non-Line-of-Sight}
 \acro{DSP}{digital signal processing}
 \acro{AFE}{analog front end}
 \acro{SQNR}{signal-to-quantisation-noise-ratio}
 \acro{ENoB}{effective number of bits}
 \acro{AGC}{automated gain control}
 \acro{PCB}{printed circuit board}
 \acro{EVM}{error vector magnitude}
 \acro{CDF}{cumulative distribution function}
 \acro{MRC}{maximum ratio combining}
 \acro{MRP}{maximum ratio precoding}
 \acro{DeepL}{deep-learning}
 \acro{DL}{downlink}
 \acro{SISO}{single input single output}
 \acro{SGD}{stochastic gradient descent}
 \acro{CP}{cyclic prefix}
 \acro{MISO}{Multiple Input Single Output}
 \acro{LMMSE}{linear minimum mean square error}
 \acro{ZF}{zero forcing}
 \acro{USRP}{universal software radio peripheral}
\end{acronym}

\begin{abstract}
We showcase the practicability of an \ac{IPS} solely based on \acp{NN} and the \ac{CSI} of a (Massive) \ac{MIMO} communication system, i.e., only build on the basis of data that is already existent in today's systems. As such our \ac{IPS} system promises both, a good accuracy without the need of any additional protocol/signaling overhead for the user localization task. In particular, we propose a tailored \ac{NN} structure with an additional \emph{phase branch} as feature extractor and (compared to previous results) a significantly reduced amount of trainable parameters, leading to a minimization of the amount of required training data. 
We provide actual measurements for indoor scenarios with up to 64 antennas covering a large area of \SI{80}{\metre^2}. 
In the second part, several robustness investigations for real-measurements are conducted, i.e., once trained, we analyze the recall accuracy over a time-period of several days. Further, we analyze the impact of pedestrians walking in-between the measurements and show that finetuning and pre-training of the \ac{NN} helps to mitigate effects of hardware drifts and alterations in the propagation environment over time. This reduces the amount of required training samples at equal precision and, thereby, decreases the effort of the costly training data acquisition.
\end{abstract}

\acresetall

\section{Introduction}
Mobile communication devices such as smartwatches and smartphones have become companions of everyday's life, resulting in a rich variety of new possible use-cases and applications in almost any area of modern life. While constant connectivity and endless computational resources have become omnipresent, the seemingly simple problem of estimating one's position inside buildings has not yet been finally resolved \cite{MSoftIndoor}.
Therefore, a need for \acp{IPS} is created, as \acp{IPS} can be seen as a key enabler for a wide range of applications such as indoor navigation, smart factories, or could even provide a basic security functionality in distributed \ac{IoT} sensor networks. 
Contrary to the outdoor scenario where \ac{GPS} provides a single universal solution for almost any possible location (assuming \ac{LoS} to the satellite), \acp{IPS} are characterized by a heterogeneous problem formulation and, thus, also many different solutions have been proposed in the literature. Moreover, also for the outdoor scenario, such a positioning system can be of practical interest as it may enhance precoding of Massive \ac{MIMO} systems through predicting the users' movements directly in the \ac{BS}. 
While the \ac{LoS} scenario is well-understood and multiple technologies are reported in the literature \cite{Iqbal2018AccurateRT,Maran2010NLOSIA,Arias2018} (e.g., angle- and time-of-arrival based predictions and triangulation methods) suitable solutions for more general channels (e.g., the much more complex \ac{NLoS} scenario) with all its practical impairments are still open for research.
Note that in the following we only focus on \ac{RF}-based technologies as it can be embedded into current systems without the need of additional sensors.

Obviously, there exists a trade-off between achievable accuracy and required overhead in terms of spectrum and computational complexity. 
In this work, we focus on rather inaccurate (targeting \emph{sub-m} precision) but low-overhead systems based on \ac{CSI} that can be implemented on top of existing communication standards. This seems to be sufficient for many applications such as indoor navigation in public buildings or movement prediction of the \ac{UE} within a \ac{BS}.
Thus, different approaches have been proposed (compare to \cite{chen2017ConFI,DeepBelief,7997235}) and investigated in the past, each optimized for different applications and system models.
Overall, these approaches can be split into two categories, where obviously mixtures between both categories exist:
\begin{enumerate}
\item Model-based: define \emph{how} the channel is expected to behave and estimate the position accordingly (e.g. ray-tracing)
\item Data-driven: collect a \emph{database} with appropriate features (often called fingerprints) and corresponding positions (e.g., CSI \cite{Khalajmehrabadi2016,Kim2018}, \ac{RSSI} \cite{bahl2000radar,savic2015fingerprintMIMO} and recently \ac{TRIPS}  \cite{wu2015TR}), i.e., somehow interpolate in-between.
\end{enumerate}
It was proposed to combine \ac{IPS} with Massive \ac{MIMO} \cite{8509634,chen2017cmIPS,vieira2017CNN,Arnold2018OnDL,Decurninge2018CSIbasedOL}, as it uses an over-provisioning of antennas to separate users in space and thereby creates an $N_{\text{Ant}}$ antenna times $N_{\text{sub}}$ subcarrier fingerprint per spatial position as a side product.  
Thus, Massive \ac{MIMO} appears to be an attractive candidate for enabling robust \acp{IPS}. 

Although there exists an underlying channel transfer function which describes the behavior of the channel for any given position, this function is typically not known or can only be approximated, as it is infeasible to fully capture the geometries of the environment and its surrounding area.
This results in the emerging of classical machine learning techniques \cite{Ghourchian2017RealTimeIL} to exploit the typical channel behavior for predicting a user's spatial position.
Moreover, it was proposed in \cite{DLSensorPos,vieira2017CNN,Arnold2018OnDL} to neglect any pre-processing which extracts channel features based on expert-knowledge, and rather directly work with raw \ac{CSI}. The intuition behind is that an \ac{NN} should be able to \emph{approximate} such functions, without the need of any \emph{a priori} knowledge other than the observed measured data. This ease in modeling and flexibility in application comes at the cost of acquiring a (typically) larger quantity of channel estimates.

It was shown in \cite{Arnold2018OnDL} that the data-driven approach can, in principle, achieve sub-cm precision, but entails an overhead of sampling large datasets per environment, and it is unclear how many samples are required to achieve a target precision.
Therefore, we investigate the influence of the amount of data samples on the prediction accuracy for standard \ac{LoS} and \ac{NLoS} indoor channels.
To improve these state-of-the-art systems, tracking over \ac{RNN} was proposed in \cite{7444919,8533813}, resulting in a more robust system. 
Although Massive \ac{MIMO} in combination with \ac{IPS} was extensively studied in \cite{8509634,chen2017cmIPS,vieira2017CNN,Arnold2018OnDL,Decurninge2018CSIbasedOL}, the influence of the number of antennas and different practical implementation challenges (e.g., hardware impairments and time varying drifts) influencing the precision and robustness of the system have yet to be studied in detail.

In addition, measurements tend to be captured over a short period of time and, thereby, often there is a lack of verifiable reproducibility of results over a longer period of time. Thus, we also investigated time dynamic effects by conducting a measurement campaign over several week days. For this, we train an \ac{NN} based on data captured at a specific day of the week, e.g. Monday, and then predicted positions based on data from another day, showing a slight performance degradation. To counter this degradation, we show that finetuning and pre-training help to mitigate those time drifting effects and also to reduce the amount of data points needed for training.
Finally, we also address another major practical burden, namely the fact that most of the time a static environment is investigated without considering any time-dependent disturbances in-between \ac{BS} and \ac{UE}, e.g., caused by pedestrians walking around as obstacles.

\vspace*{-0.2cm}
\section{Background}\label{sec:model_and_limits}
\vspace*{-0.15cm}
It is widely accepted that Massive \ac{MIMO} unleashes its full potential only in combination with \ac{TDD} as the piloting overhead is independent of the number of \ac{BS} antennas $N_\text{Ant}$  \cite{EmilBjoernson2017}. For simplicity, we assume that a \ac{UE} is only equipped with a single antenna, but extensions are straightforward. Further, we use \ac{OFDM}, i.e., $N_{\text{sub}}$ subcarriers exist.

\begin{figure}
    \centering
    \includegraphics{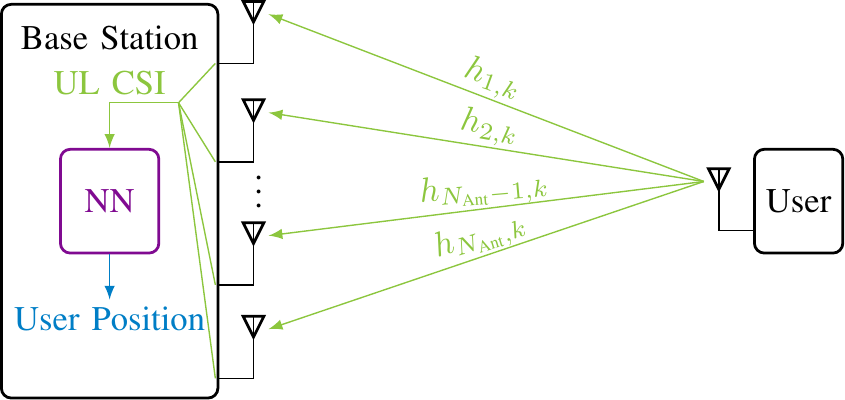}
    \caption{System schematic for predicting a user's position with an \ac{NN} based on \ac{UL} \ac{CSI}.} 
    \label{fig:System-Model-Chan-Est}
    \vspace*{-0.4cm}
\end{figure}

For standard communication in \ac{TDD} Massive \ac{MIMO} the \acp{UE} transmit orthogonal pilots to the \ac{BS} and 
the \ac{BS} uses these pilots to estimate the channel at antenna $m$ $\hat{\hv}_{m,k} \in \mathbb{C}^{N_\text{Ant} \times N_\text{sub}}$ per subcarrier $k$ as shown in Fig.~\ref{fig:System-Model-Chan-Est}. 
This estimate of the channel (\ac{CSI}) is then used to orthogonalize the users in space.
In the following, we reuse this \ac{CSI} to create a robust fingerprint for each spatial position $x,y,z$ of the \ac{UE}.

To compare the prediction performance we use two different metrics:
\begin{enumerate}
\item The \ac{MDE} \begin{equation}
\ac{MDE} = \frac{1}{N_{\text{test}}}\sum^{N_{\text{test}}}_{n=1}  \Vert \dv_n-\hat{\dv_n}  \Vert _2
\end{equation}
is used for simulated data, where the average \ac{DE} is calculated over the whole test size $N_{\text{test}}$.
\item Since this metric punishes outliers, we use the \ac{MDA} for measured data (compare \cite{DLSensorPos}).
\ac{MDA} is defined as the \ac{DE} that 50\% of the users are achieving at least, which thereby removes the influence of heavy outliers that often occur in measurements.
\end{enumerate}

\subsection{Simulated Channels}

\begin{figure}[H]
    \centering
    \includegraphics{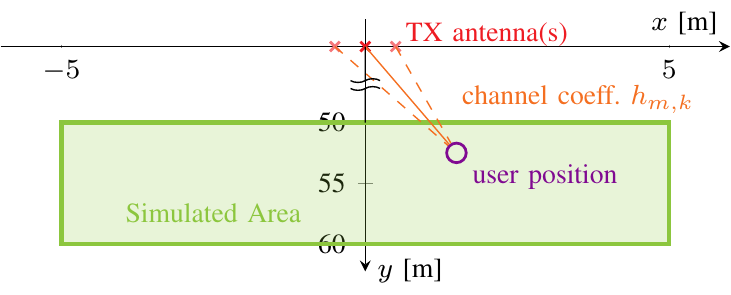}
    \vspace*{-0.2cm}
    \caption{Simulation setup for the \emph{3GPP} channel model.} %\comsc{kleiner; auf y-achse noch unterbrechungen einfügen}
    \label{fig:SimArea}
    \vspace*{-0.4cm}
\end{figure}

To show the viability of the proposed \ac{NN}, we begin with an investigation on simulated standardized channel models for \ac{LoS} and \ac{NLoS} indoor scenarios.
For this, we use the Quadriga Framework \cite{Jaeckel2014}.
Fig.~\ref{fig:SimArea} illustrates the simulation setup, where a \ac{BS} with the antenna geometry of an $8\times8$ patch array is placed in the origin of the coordinates and an area of \SI{100}{\metre^2} is simulated.
The sampling resolution is \SI{10}{\centi \metre} $< \lambda/2$ in both $x$ and $y$ direction.
Further, an \ac{OFDM} channel estimation with 1024 subcarriers and a bandwidth of \SI{20}{\mega \hertz} around the carrier frequency of \SI{1.25}{\giga \hertz} is used.
These parameters are chosen to equal those of the actual measurement campaigns as presented in the later sections.

\subsection{Measured Channels}
\begin{figure}[H]
     \centering
    \includegraphics{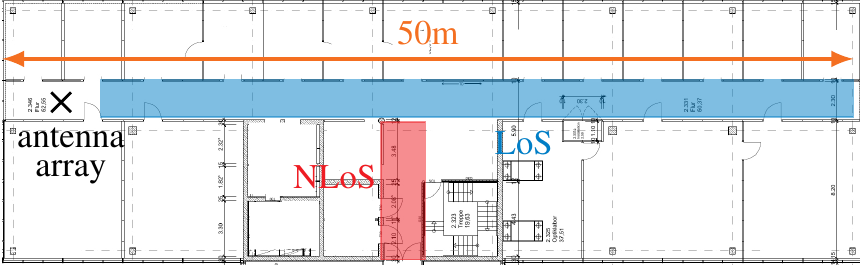}
     \caption{Indoor measurement positions.}
     \label{fig:MeasurementScenarios}
\end{figure}

We provide a short introduction to the actual measurements that were conducted at our institute, as can be seen in Fig.~\ref{fig:MeasurementScenarios}. For the exact measurement procedure which inherently provides ground truth (i.e., 3D position labels) at \emph{centimeter} precision, we refer the interested reader to \cite{Arnold2018}.

The transmitter (\ac{UE}) consists of an amplified \ac{USRP} with a dipole antenna which transmits \ac{OFDM} pilot symbols with 1024 subcarriers and a bandwidth of \SI{20}{\mega \hertz} at \SI{1.25}{\giga \hertz}.
For subcarrier modulation, a simple \ac{BPSK} constellation was chosen.
Note that 10\% of the subcarriers are used as guard band and the cyclic prefix was 1/8 of the \ac{OFDM} symbol duration.
To be able to reuse the same \ac{NN} structure for simulated and measured samples, the missing guard band subcarriers in measured samples were replaced by zeros to maintain the \ac{NN}'s input size of $1024$ subcarriers.

\begin{table}[H]
    \centering
    \caption{Description of datasets.}
    \begin{tabular}{c|c|c}
        Dataset             & $\#$ Samples  & Covered Area       \\ \hline
        Indoor Simulations                 & 10000                &  \SI{10}{\metre} $\times$ \SI{10}{\metre}=\SI{100}{\metre^2}        \\
        \ac{LoS} Weekdays Measurements             & 6000                 &  \SI{40}{\metre} $\times$ \SI{2}{\metre}=\SI{80}{\metre^2}                    \\
        Disturbed Indoor Measurement          & 5800               &  \SI{40}{\metre} $\times$ \SI{2}{\metre}=\SI{80}{\metre^2}                    \\
        \ac{NLoS} Indoor Measurement                   & 2700               &  \SI{2}{\metre} $\times$ \SI{18}{\metre}=\SI{36}{\metre^2}                        \\
    \end{tabular}
    \label{tab:Datasets}
    \vspace*{-0.4cm}
\end{table}

Tab.~\ref{tab:Datasets} gives an overview of the used datasets and their corresponding coverage areas, where the area dimensions match those in \cite{DLSensorPos}.
In the ``\ac{LoS} Weekdays Measurements'' each day was measured with the same meander-like path structure, resulting in a sample distance of around \SI{1}{\centi \metre}.
For the ``Disturbed Indoor Measurement'', two colleagues were shadowing the antenna array by randomly walking in-between \ac{BS} and \ac{UE}.

\section{Proposed Neural Network Architecture}
\begin{figure}
   \vspace{-0.2cm}
    \centering
    \includegraphics{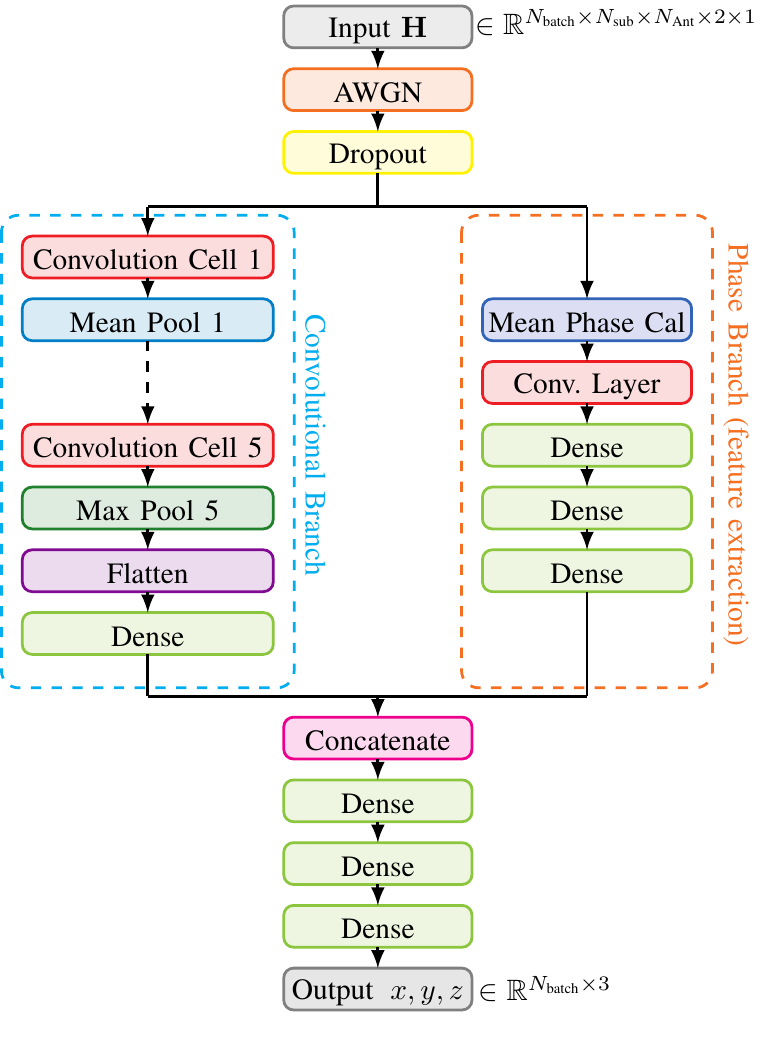}
     \vspace{-0.5cm}
    \caption{Basic structure of the proposed \ac{NN}.}
    \label{fig:DLNNStruct}
         \vspace{-0.4cm}
\end{figure}
Fig.~\ref{fig:DLNNStruct} depicts the layout of the proposed \ac{NN}.
As we treat complex values as two independent real numbers, the input of the \ac{NN} has the shape
\begin{equation*}
N_{\text{batch}} \times N_{\text{sub}} \times N_{\text{Ant}} \times 2.
\end{equation*}
where $ N_{\text{sub}} $ is the number of subcarriers, $ N_{\text{Ant}} $ is the number of antennas, and the fourth dimension is composed of the real and imaginary parts.
A noise and a dropout layer is directly added to this input to prevent the \ac{NN} from overfitting and to reinforce the \ac{NN} not to rely on strong antennas only.
Afterward, the graph is split in two branches, a ``Convolutional Branch'' and a ``Phase Branch''.
The ``Convolutional Branch'' is built of $5$ convolutional cells for evaluating the fingerprint in the amplitude of the real and imaginary part.
Therefore, a 3-dimensional kernel of shape $X \times Y \times 2$ is used in the first convolutional layer, to combine real and imaginary parts, while subsequent layers use 2-dimensional kernels to further convolve over the subcarrier and antenna dimensions.
The second branch, referred to as ``Phase Branch'', further improves the \ac{MDE} (simulated data)/ \ac{MDA} (measured data), by focusing on the phase difference between antennas, which helps especially when measured data is used.
Within this branch we explicitly calculate the mean phase per antenna over all subcarriers and then forward this refined information to the \ac{NN}.
Inserting such expert knowledge operations drastically reduces complexity that otherwise must be learned through exhaustive training. This is similar to the idea of \emph{transformer networks} \cite{oshea2017introduction}. By doing so, the \ac{NN} experiences a faster convergence and a higher final prediction accuracy.
Finally, the two branches are concatenated and four fully connected dense layers are used to combine both outputs to arrive at a prediction of the user's $x-,y-,z$-position.

During training, we use $N_\text{train}$ samples (i.e., data $\mathbf{v}$ and position labels $\mathbf{w}$) and apply multiple stochastic gradient descent iterations, where one iteration over the whole dataset is denoted as\emph{epochs} $N_\text{ep}$.
Thus, the \emph{same} $N_\text{train}$ samples are considered $N_\text{ep}$ times, to find the best weights $\boldsymbol{\theta}$.
We chose the Huber-loss \cite{10.2307/2238020} as loss metric during training and the absolute distance deviation to measure the accuracy.
Intuitively, the optimal training \ac{SNR} is a trade-off between high noise power, i.e,. \emph{learning robustness to noisy data} and noiseless samples, i.e., \emph{learning the underlying (deterministic) channel transfer function} \cite{gruber2017}.
To further prevent overfitting, we used the previously mentioned dropout layer where we reached best performance and accuracy with a dropout rate of $ 10 \si{\percent} $.
Additionally, the generalization effect of the dropout layer comes along with an improved prediction with respect to reproducibility in time for changing datasets, as will be shown later.
We start the training procedure with mini-batch sizes of $ 16 $ samples.
After retraining for multiple epochs until an early stopping mechanism detects no further improvements, we then continue training with a stepwise increased batch size up to $ 512 $ samples per batch.
To achieve a higher final prediction accuracy, the learning rate is also reduced stepwise from $0.001$ to $ 0.00005$ before advancing to the next batch size.

\begin{table}[H]
    \centering
    \caption{Improvement by using a phase branch}
    \begin{tabular}{c|c|c|c}
        Dataset             & \multicolumn{1}{ c| }{without}  & \multicolumn{1}{ c| }{Phasebranch}     & \multicolumn{1}{ c }{Rel. Impr. }        \\ \hline
        3GPP \acs{LoS}       & \SI{0.1621}{\meter}                &  \SI{0.0863}{\meter}               &  \SI{46.1}{\percent} \\
        3GPP \acs{NLoS}        & \SI{0.1275}{\meter}                &  \SI{0.1183}{\meter}                     &  \SI{7.2}{\percent} \\
        Monday Mes.           & \SI{0.3467}{\meter}                &  \SI{0.2460}{\meter}                &  \SI{29.0}{\percent} \\
    \end{tabular}
    \label{tab:Phasebranch}
\end{table}
Tab.~\ref{tab:Phasebranch} shows the gain of the proposed ``Phase Branch'' over traditional fingerprinting approaches, resulting in an improvement in measurements of \SI{29}{\percent} by only adding $22000$ weights.
It can be seen that for the simulated data and measured data the \ac{NN} achieves (with a total of only 440.000 weights) the same accuracy as in \cite{DLSensorPos} for a similar area and propagation environment (cf. Fig. in \cite{DLSensorPos}).

\subsection{Required Training Data}\label{sec:SimSparse}
In a practical use-case, training data is valuable as each sample has to be captured manually and also the ground truth position is required. Therefore, we investigate how the \ac{NN} approach can deal with a larger grid resolution than the original one (\SI{10}{\centi\meter}) with 10.000 samples.
To make a fair comparison we take every second/third/.. sample, in the mesh grid, i.e., create a larger sample distance. Note that the sample size is reduced quadratically (both dimensions $x$ and $y$).
\begin{figure}[H]
    \vspace*{-0.3cm}
    \centering
    \includegraphics{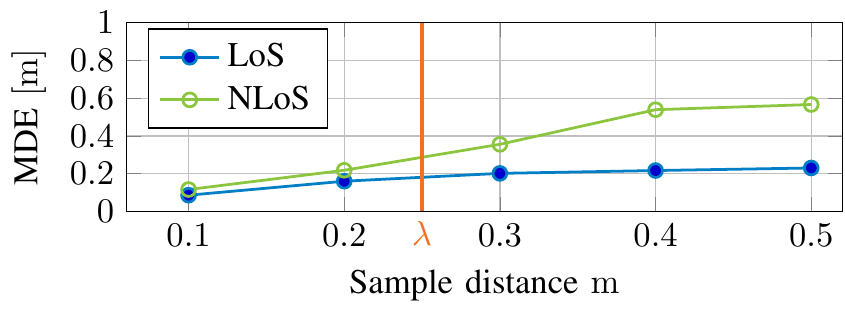}
    \vspace*{-0.3cm}
    \caption{Simulated: 3GPP \ac{LoS} and \ac{NLoS} outdoor \ac{MDE} for different sample distances.}
    \label{fig:SparseInv}
\end{figure}
Fig. ~\ref{fig:SparseInv} shows the \ac{MDE} for the different sample distances. In the \ac{LoS} case the \ac{NN} is able to resolve the positions even with a low grid resolution. The \ac{NLoS} is far more dependent on a finer grid resolution, as it has less spatial correlation and inhibits more random components than the \ac{LoS} case. Therefore the \ac{NN} is less capable of interpolating in between the trained samples.
From this it can be seen, that in the \ac{LoS} case the \ac{NN} learns a more robust representation of the \ac{LoS} fingerprints than in the \ac{NLoS} case.

\subsection{Influence of the Number of Antennas}

In typical Massive \ac{MIMO} scenarios the antenna gain and the ability to separate users increases with the number of antennas, therefore can be expected that by increasing the number of antennas the fingerprint and the overall system is more robust against different impairments. However, also the training complexity increases.
\begin{figure}[H]	
    \centering
    \includegraphics{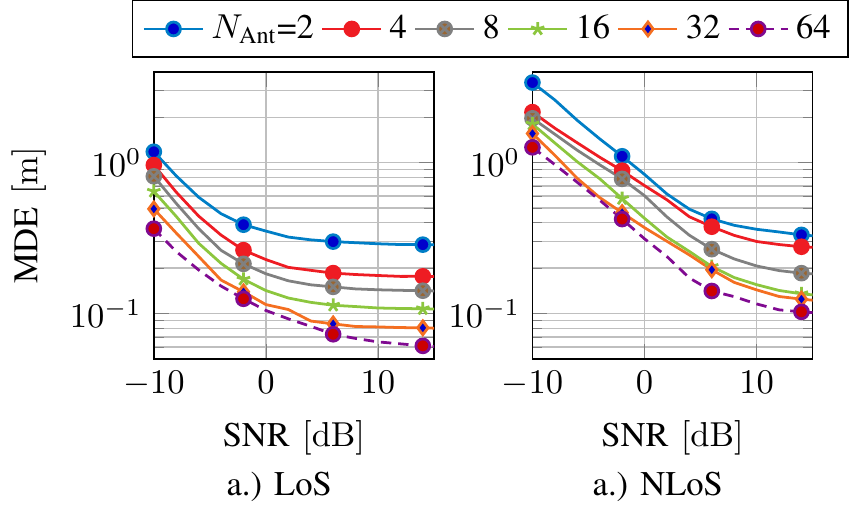}
    \vspace*{-0.3cm}
    \caption{Simulated: \ac{MDE} for the 3GPP \ac{LoS} indoor channel models with different number of antennas, trained at a target $\ac{SNR}=\SI{20}{\decibel}$.}
    \label{fig:3GPP_LOS_Antsweep}
\end{figure}
Fig.~\ref{fig:3GPP_LOS_Antsweep} illustrates the robustness of the proposed system for different number of antennas in the \ac{LoS} case (left) and in the \ac{NLoS} case (right). It can be seen that roughly a \SI{3}{\decibel} gain per doubling the antenna occurs. But even with a low number of antennas 
the system still achieves reasonably good results in the case of \ac{LoS}. For the \ac{NLoS} case, more antennas are needed for achieving a robustness against measurement noise (curves are shifted to the right) and more antennas are needed to achieve similar accuracy as in the \ac{LoS} case. This shows that the number of antennas enhances the robustness in all cases for \ac{IPS} and that Massive \ac{MIMO} is a good match for this system.
\begin{figure}[H]
    \centering
    \includegraphics{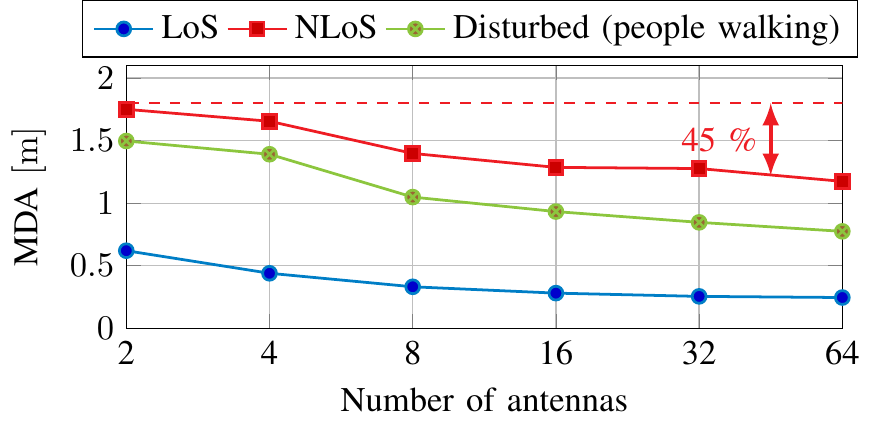}
    \vspace*{-0.5cm}
    \caption{Measured: \ac{MDA} with different numbers of antennas for measured \ac{NLoS}, \ac{LoS} and disturbed \ac{LoS}  datasets.}
    \label{fig:Mes_Antsweep}
    \vspace*{-0.4cm}
\end{figure}
Fig.~\ref{fig:Mes_Antsweep} depicts for the dependency of the \ac{MDA} on the different amount of antennas. It can be seen that here the \ac{LoS} case shows only a small improvement by using $64$ antennas instead of 8. In contrast, the robustness of the whole system for the \ac{NLoS} and ``Disturbed Indoor Measurement'' case can be significantly improved (over 50\% performance gain). Therefore, the same behavior as the simulated case is exhibited. Although in simulations this system could be used with a smaller number of antennas, the effect of hardware impairments and disturbance through the movements/obstacles in the measurement area is reduced by increasing the number of antennas.
As those measurements were conducted only within a fixed time interval, an important question remains regarding the overhead to achieve the same performance if the environment changes (e.g. movement) and/or when the hardware impairments become significant.

\section{Robustness and time Reproducibility}\label{sec:MeasTime}
At first we investigate the performance loss due to changing propagation environments as can be caused by moving obstacles, such as pedestrians or cars.
As the main focus of this work is the indoor scenario, we consider ``indoor pedestrians'' randomly walking through the measurement area.
It is important to realize that such obstacles do not only cause a simple signal attenuation, but also may change the propagation scenario from \ac{LoS} characteristics to \ac{NLoS} behavior or even shadow several antennas.
\begin{figure}[H]
    \centering
    \includegraphics{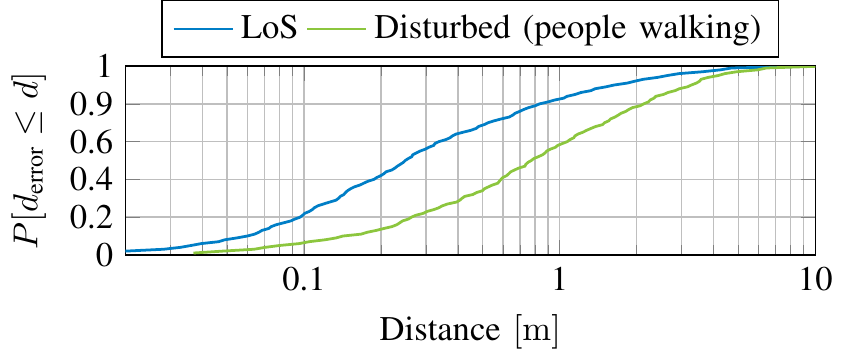}
    \vspace*{-0.3cm}
    \caption{Measured: \ac{DE} \ac{CDF} for pure \ac{LoS} and disturbed \ac{LoS} measurements on the same area.}
    \label{fig:TimeInv_percentil}
\end{figure}
Fig.~\ref{fig:TimeInv_percentil} shows the \ac{DE} \acf{CDF} of different investigations into time reproducibility. For the perfect \ac{LoS} case, the system achieves a quite reasonable precision in the range of \SI{23}{\centi \metre} on an \SI{80}{\metre^2} area. In contrast to the static scenario, in the ``Disturbed Indoor Measurement'' scenario the performance decreases to around \SI{70}{\centi \metre}, which is still sufficiently accurate for many practical applications.

\subsection{Reproducibility Over Time}
Another important property of a practical \ac{IPS} is the reproducibility over time, i.e., once trained it needs to provide a stable accuracy over time.
\begin{figure}
\vspace*{-0.3cm}
\centering
    \includegraphics{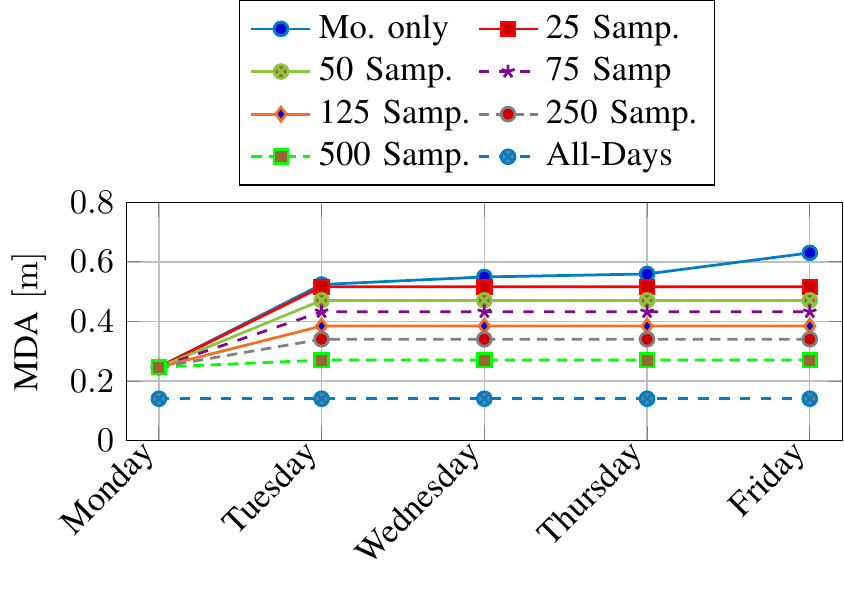}
    \vspace*{-0.3cm}    
    \caption{Measured: Enhancing time reproducibility via finetuning with different amounts of samples.}
    \label{fig:TimeInv2}
    \vspace*{-0.3cm}
\end{figure}
Fig.~\ref{fig:TimeInv2} shows the effect of retraining the \ac{NN} (that was initially trained on Monday) each day with a different number of points.
Although the hardware was turned off and on between the measurement days and the propagation environment may have slightly changed (e.g., due to open doors and windows), it can be seen that even without finetuning the system is still able to achieve a relatively good accuracy of \SI{55}{\centi \metre} when inferred on other days.
This shows the reproducibility over several time incoherent measurements.
To further improve the performance, we propose to measure a small amount of ``calibration'' points and perform finetuning only on these few points. It can be seen that with 125 samples per day, the loss can be significantly reduced and an accuracy of \SI{40}{\centi \metre} is reached. In conclusion, this shows that the proposed \ac{NN}-based user positioning system is robust to obstacles and, once trained, remains stable over time. 

\subsection{Pre-training with Simulated Data}
We now aim at lowering the amount of required training data and investigate the effects of pre-training.
In particular, we compare three different methods:
\begin{enumerate}
\item Random initialization of weigths (e.g. Xavier initalization \cite{pmlr-v9-glorot10a}).
\item Pre-training based on the simulated \emph{3GPP} model.
\item Pre-training based on a subset of subcarriers, resulting in data augmentation
\end{enumerate}
The intuition behind using only a subset of subcarriers for the initial training is the same as in \cite{Perez2017TheEO} (data augmentation), where a cropped or rotated image is used to virtually extend the training dataset with similar data
in just another presentation.
This approach also resembles dropout and results in requiring less samples for achieving an even better overall performance \cite{Perez2017TheEO}.
\begin{figure}
    \centering
    \includegraphics{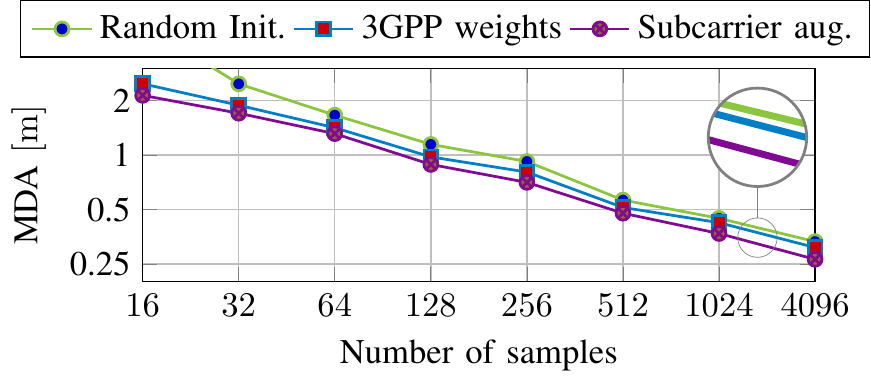}
    \vspace*{-0.5cm}
    \caption{Measured: Achieved \ac{MDA} on Monday dataset for different initialization methods and sample distances.} 
    \label{fig:Histogram}
    \vspace*{-0.4cm}
\end{figure}
Fig.~\ref{fig:Histogram} shows the \ac{MDA} of the pre-trained model trained on a new propagation environment for different numbers of sample sizes.
As expected, the \ac{MDA} decreases with increasing sample size for all investigated methods.
However, even at a high number of samples, there is still a gain of about \SI{10}{\centi \metre} for the proposed methods over a randomly initialized \ac{NN}.
We want to emphasize that all data samples obtained by the simulated model or through data augmentation can be considered as free of cost, in contrary to actual measurement data.
The \ac{NN} is able to learn the new scenario with fewer training samples in these pre-trained cases, as the \ac{NN} can refine its weights according to the propagation environment.

\section{Conclusions and Outlook}\label{sec:conclusions}

We have shown the practical viability of \ac{IPS} based on \ac{CSI} of a Massive \ac{MIMO} systems even in measured \ac{NLoS} scenarios with complex propagation environments where most existing solutions would fail. Therefore, a novel \ac{NN} structure has been proposed based on the idea of an additional feature extraction branch (\emph{phase branch}), which turned out to improve the performance by a great margin. We showed the robustness of the system for both, moving obstacles in the measurement area as well as in terms of reproducibility over time, i.e., we showcased that, once trained, the system maintains reasonable accuracy over many days. Further, we proposed finetuning and pre-training of the \ac{NN} to mitigate the effects of varying hardware impairments and changes in the propagation environment which turns out to reduce the number of data points needed for training, resulting in a reduction of the high cost of capturing precise training points.

\bibliographystyle{IEEEtran}
\bibliography{IEEEabrv,references}

\end{document}